\documentclass[a4paper,11pt]{article}
\usepackage[latin1]{inputenc}
\usepackage[T1]{fontenc}
\usepackage[english]{babel}
\usepackage{graphicx}
\usepackage{epsf}
\begin{document}
\title{Isotropisation of flat homogeneous Bianchi type $I$ model with a non minimally coupled and massive scalar field}
\author{Stéphane Fay\footnote{Laboratoire Univers et Théories, CNRS-UMR 8102 - Observatoire de Paris, F-92195 Meudon Cedex - France}\footnote{Steph.Fay@Wanadoo.fr}
}
\maketitle
\begin{center}
\begin{abstract}
In previous works, we studied the isotropisation of Bianchi class $A$ models with a minimally coupled scalar field. In this paper, we extend these results to the case of a non minimally coupled one. We first make the calculations in the Einstein frame where the scalar field is minimally coupled to the curvature but non minimally coupled to the perfect fluid. Then, we use a conformal transformation to generalise our results to a scalar field non minimally coupled to the curvature. Universe isotropisation for the Brans-Dicke and low energy string theories are studied.
\end{abstract}
\end{center}
\section{Introduction}\label{s0}
There are numerous reasons to consider the presence of some scalar fields in our Universe. Historically, the most famous scalar tensor theory is the Brans-Dicke one which aimed to satisfy Mach ideas. Since the eighties some new justifications appeared mainly related to particle physics theories. As instance, supersymmetry supposes the equality between fermionic and bosonic degrees of freedom and needs several scalar fields to exist. The Higgs field is a scalar field. Some other reasons to believe that scalar fields exist are related to the cosmology: they could explain the flattening of some spiral galaxies rotation curves\cite{MatGuzNum00, Fay03A}(dark matter), the late time Universe accelerating expansion\cite{Per99, Rie98}(dark energy) or the inflation.\\
In this work we will be interested in a massive scalar field, non minimally coupled to the curvature. In \cite{TorVuc96}, it is called the Hyperextended Scalar Tensor (HST). It is such that the gravitationnal constant varies with respect to the scalar field. We will perform the calculations in the Einstein frame for which, after a conformal transformation of the metric, the field is cast into a minimally coupled and massive one. However it is then coupled to the perfect fluid. Consequently the matter does not follow the spacetime geodesics.\\
The geometrical framework will be the homogeneous and spatially flat Bianchi type $I$ model. It is an anisotropic model which generalises the flat Friedmann-Lemaître-Robertson-Walker (FLRW) one. The Bianchi models allows understanding how the isotropisation appeared. If the isotropy and homogeneity of our Universe is well established until the decoupling period\cite{Spe03}, one has to remember that it is a hypothesis concerning the early Universe. Another justification for an initially anisotropic state of the Universe is that the FLRW singularity approach is not generic. A generic approach could be oscillating as the one of the Bianchi type $IX$ model. It has been conjectured by Belinskij, Khalatnikov and Lifchitz (BKL)\cite{BelKhaLif82,BelKhaLif70} that it should be shared by the most general anisotropic and inhomogeneous models. This conjecture has been revisited recently by Uggla and others\cite{UggElsWaiEll03}. Here, we will study the Bianchi type $I$ model whose singularity is not oscillatory but which is a spatially flat model in agreement with WMAP data.\\
Our goal will be to study the isotropisation process of the Bianchi type $I$ model in presence of a non minimally coupled and massive scalar field. To this end, we will use the ADM Hamiltonian formalism\cite{Nar72,Mis62}. It allows writing the field equations as a first order differential system. We will use the dynamical systems methods\cite{WaiEll97} to find its isotropic equilibrium states. The plan of the paper is as follows. In section \ref{s1}, we write the Hamiltonian field equations. In section \ref{s2}, some necessary conditions for isotropisation and the behaviours of the metric and potential when such a state is reached are described. In the last section, we summarize our results and study the isotropisation of the Brans-Dicke and low energy string theories when the potential is a power or an exponential law of the scalar field.
%------------------------------------------------------------------------------------------------------------------------------------------------------------------------------------------%
\section{Field equations}\label{s1}
The action of the HST in the Einstein frame writes:
%----------------------------------EQUATION----------------------------------------%
\begin{equation} \label{action}
S=(16\pi)^{-1}\int \left[R-(3/2+\omega)\phi^{,\mu}\phi_{,\mu}\phi^{-2} -U\right]\sqrt{-g}d^4 x+S_m(g_{ij},\phi)
\end{equation}
$\phi$ is the scalar field, $\omega$ and $U$ are respectively the Brans-Dicke coupling function and the potential. They depend on $\phi$. $S_m$ is the action standing for a perfect fluid coupled to the scalar field. Its equation of state is $p_m=(\gamma-1)\rho_m$ with $\gamma\in\left[1,2\right]$. The metric for the Bianchi type $I$ model is:
%-------------------------EQUATION-------------------------------------%
\begin{equation}
ds^2=-dt^2+R_0 ^2 g_{ij}\omega^i\omega^j
\end{equation}
The $g_{ij}$ are the metric functions and $\omega_i$ the 1-forms specifying the Bianchi type $I$ model. In order to use the Hamiltonian formalism, we rewrite this metric with a 3+1 decomposition of spacetime:
\begin{equation}
ds^2 = -(N^2 -N_i N^i )d\Omega^2 + 2N_i d\Omega\omega^i + R_0 ^2 e^{-2\Omega+2\beta_{ij}}\omega^i \omega^j 
\end{equation}
$N$ and $N_i$ are the lapse and shift functions. $\Omega$ describes the isotropic part of the metric and will be considered as a time coordinate. We will show latter that it is a monotonic function of the proper time $t$. Moreover, we define a comoving 3-volume $V$ as $V=e^{-3\Omega}$. The $\beta_{ij}$ stand for the anisotropic part of the metric. they were parameterised by Misner\cite{Mis62} in the following way:
%-----------------------EQUATION-------------------------------------%
\begin{equation}
\beta_{ij}=diag(\beta_++\sqrt{3}\beta_-,\beta_+-\sqrt{3}\beta_-,-2\beta_+)
\end{equation}
%-----------------------EQUATION-------------------------------------
\begin{equation}
p_k^i=2\pi\pi_k^i-2/3\pi\delta_k^i\pi_l^l
\end{equation}
%-----------------------EQUATION-------------------------------------%
\begin{equation}
6p_{ij}=diag(p_++\sqrt{3}p_-,p_+-\sqrt{3}p_-,-2p_+)
\end{equation}
The $p_{ij}$ are the $\beta_{ij}$ conjugate momenta. To find the ADM Hamiltonian, we rewrite the action as:
%-----------------------EQUATION-------------------------------------%
\begin{equation}\label{action1}
S=(16\pi)^{-1}\int(\pi^{ij}\frac{\partial{g_{ij}}}{\partial{t}}+\pi^{\phi}\frac{\partial{\phi}}{\partial{t}}+\pi^{\psi}\frac{\partial{\psi}}{\partial{t}}-NC^0-N_iC^i)d^4x
\end{equation}
$\pi_\phi$ is the scalar field conjugate momentum. $N$ and $N_i$ are similar to some Lagrange multipliers. By varying the action with respect to these quantities, we find the constraints $C_0=0$ and $C_i=0$ with:
%----------------------------------EQUATION----------------------------------------%
\begin{eqnarray*}
C^0&=&-\sqrt{^{(3)}g}^{(3)}R-\frac{1}{\sqrt{^{(3)}g}}(\frac{1}{2}(\pi^k _k )^2 -\pi^{ij}\pi_{ij})+\frac{1}{2\sqrt{^{(3)}g}}\frac{\pi_\phi ^2 \phi^2 }{3+2\omega}+\nonumber\\
&&\sqrt{^{(3)}g}U+\frac{1}{\sqrt{^{(3)}g}}\frac{\delta \lambda e^{3(\gamma-2)\Omega}}{24\pi^2}\nonumber\\
C^i&=&\pi^{ij}_{\mid j}
\end{eqnarray*}
$\delta$ and $\lambda$ are respectively a positive constant and a scalar field function describing the coupling between the scalar field and the perfect fluid. The action (\ref{action}) may be derived from the one of a non minimally coupled scalar field as described in the appendice. Then the gravitation function $G(\phi)$ is variable and we have the relation $\lambda\propto G^{3(4-3\gamma)}$. The energy conservation of the perfect fluid writes $\rho_m=\lambda V^{-\gamma}$. Hence, we will assume that $\lambda$ is a positive function of $\phi$. The constraint $C_i=0$ is identically satisfied whereas the constraint $C_0=0$ gives the ADM Hamiltonian:
%----------------------------------EQUATION----------------------------------------%
\begin{equation} \label{hamiltonian}
H^2 = p_+ ^2 +p_- ^2 +12\frac{p_\phi ^2 \phi^2}{3+2\omega}+24\pi^2 R_0 ^6 e^{-6\Omega}U+\delta \lambda e^{3(\gamma-2)\Omega}
\end{equation}
When $\lambda=const$, we recover the Hamiltonian when the scalar field is not coupled to the perfect fluid. For more details on Hamiltonian formalism, see \cite{Rya72}. The Hamiltonian equations then write:
%----------------------------------EQUATION----------------------------------------%
\begin{equation} \label{betapm}
\dot{\beta}_ \pm = \frac{\partial H}{\partial p_ \pm}=\frac{p_\pm}{H}
\end{equation}
%----------------------------------EQUATION----------------------------------------%
\begin{equation} \label{phi}
\dot{\phi}=\frac{\partial H}{\partial p_\phi}=\frac{12\phi^2 p_\phi }{(3+2\omega)H}
\end{equation}
%----------------------------------EQUATION----------------------------------------%
\begin{equation} \label{ppm}
\dot{p}_\pm=-\frac{\partial H}{\partial \beta_ \pm}=0
\end{equation}
%----------------------------------EQUATION----------------------------------------%
\begin{equation} \label{pphi}
\dot{p}_\phi=-\frac{\partial H}{\partial \phi}=-12\frac{\phi p_\phi ^2}{(3+2\omega)H}+12\frac{\omega_\phi \phi^2 p_\phi ^2 }{(3+2\omega)^2 H}-12\pi^2 R_0 ^6 \frac{e^{-6\Omega}U_\phi }{H}-\frac{\delta \lambda_\phi e^{3(\gamma-2)\Omega}}{2H}
\end{equation}
%----------------------------------EQUATION----------------------------------------%
\begin{equation} \label{H}
\dot{H}=\frac{dH}{d\Omega}=\frac{\partial H}{\partial \Omega}=-72\pi^2 R_0 ^6 \frac{e^{-6\Omega}U}{H}+3/2\delta\lambda(\gamma-2)\frac{e^{3(\gamma-2)\Omega}}{H}
\end{equation}
A dot means a derivative with respect to $\Omega$ and a subscript $\phi$ a derivative with respect to the scalar field. We rewrite these equations with some bounded variables. For this we define:
%-------------------------EQUATION-------------------------------------%
\begin{equation} \label{var1}
x=H^{-1}
\end{equation}
%-------------------------EQUATION-------------------------------------%
\begin{equation} \label{var2}
y= e^{-3\Omega}\sqrt{U}H^{-1}
\end{equation}
%-------------------------EQUATION-------------------------------------%
\begin{equation} \label{var3}
z=p_{\phi}\phi(3+2\omega)^{-1/2}H^{-1}
\end{equation}
These new variables will be real if $U>0$ and $3+2\omega>0$. Thus the scalar field respects the weak energy condition. Each variable have a physical interpretation:
\begin{itemize}
\item $x^2$ is proportional to the shear parameter $\Sigma$ defined in \cite{WaiEll97}.
\item $y^2$ is proportional to $(\rho_\phi-p_\phi)/(d\Omega/dt)^2$, $(d\Omega/dt)^2$ being the Hubble constant when the Universe is isotropic, $\rho_\phi$ and $p_\phi$ the density and pressure of the scalar field.
\item $z^2$ is proportional to $(\rho_\phi+p_\phi)/(d\Omega/dt)^2$.
\item From the two last points it comes that the density parameter of the scalar field, $\Omega_\phi\propto \rho_\phi/(d\Omega/dt)^2$, is a linear combination of $y^2$ and $z^2$. When the scalar field is quintessent these two variables are proportional to $\Omega_\phi$.
\end{itemize}
From the equation (\ref{hamiltonian}), we get:
%-------------------------EQUATION-------------------------------------%
\begin{equation} \label{11}
p^2x^2+R^2y^2+12z^2+k^2=1
\end{equation}
where we put to simplify 
$$
k^2=\delta\lambda e^{3(\gamma-2)\Omega}H^{-2}
$$
and we define the constants $p^2=p_+^2+p_-^2$ and $R^2=24\pi^2R_0^6$. The equation (\ref{11}) may be considered as a constraint equation. It shows that the variables $x$, $y$, $z$ and $k$ are bounded. $k$ is not a new independent variable but is related to $x$, $y$ and $\phi$, this last variable being able to diverge. It is proportional to the perfect fluid density parameter $\Omega_m\propto \rho_m/(d\Omega/dt)^2$ and it may be rewritten under the following useful forms:
%-------------------------EQUATION-------------------------------------%
\begin{equation}
k^2=\delta\lambda x^\gamma y^{2-\gamma}U^{\gamma/2-1}\nonumber
\end{equation}
%-------------------------EQUATION-------------------------------------%
\begin{equation}\label{form1}
k^2=\delta\lambda x^2e^{3(\gamma-2)\Omega}
\end{equation}
%-------------------------EQUATION-------------------------------------%
\begin{equation}\label{form2}
k^2=\delta y^2 U^{-1} \lambda V^{-\gamma}
\end{equation}
Using the variables (\ref{var1}-\ref{var3}), the field equations become:
%-------------------------EQUATION-------------------------------------%
\begin{equation} \label{12}
\dot{x}=3R^2y^2x-3/2(\gamma-2)k^2x
\end{equation}
%-------------------------EQUATION-------------------------------------%
\begin{equation} \label{13}
\dot{y}=y(6\ell z+3R^2y^2-3)-3/2(\gamma-2)k^2y
\end{equation}
%-------------------------EQUATION-------------------------------------%
\begin{equation} \label{14}
\dot{z}=R^2y^2(3z-\frac{\ell}{2})-3/2(\gamma-2)k^2z-1/2\ell_mk^2
\end{equation} 
where the quantities $\ell$ and $\ell_m$ are defined by $\ell=\phi U_\phi U^{-1}(3+2\omega)^{-1/2}$ and $\ell_m=\phi \lambda_\phi \lambda^{-1}(3+2\omega)^{-1/2}$. $\ell$ and $\ell_m$ look each other because of the similar roles of $U$ and $\lambda$ in the Hamiltonian (\ref{hamiltonian}). Both are multiplied by an exponential of $\Omega$. The equation for $\phi$ will be written as:
%-------------------------EQUATION-------------------------------------%
\begin{equation}\label{15}
\dot{\phi}=12z\frac{\phi}{(3+2\omega)^{1/2}}
\end{equation}
Summarising, the seven equations of the Hamiltonian system (\ref{betapm}-\ref{H}) are reduced to a system of four equations (\ref{12}-\ref{15}). It describes the evolution of four variables of which three are bounded. It comes owing to the fact that, for the Bianchi type $I$ model, the hamiltonian equations immediately give $p_\pm =const$ implying $\beta_+\propto\beta_-$. Moreover, we will choose a diagonal form for the metric, i.e. $N_i=0$. It allows  getting $N=12\pi R_0^3H^{-1}e^{-3\Omega}$ with $dt=-Nd\Omega$. 
%------------------------------------------------------------------------------------------------------------------------------------------------------------------------------------------%
\section{Stable isotropic states}\label{s2}
%------------------------------------------------------------------------------------------------------------------------------------------------------------------------------------------%
\subsection{Defining isotropy}\label{s21}
Following Collins and Hawking\cite{ColHaw73}, isotropy arises when $\Omega\rightarrow -\infty$. Universe is thus forever expanding Universe with $d\beta_\pm/dt\propto e^{3\Omega}\rightarrow 0$. Moreover, defining $\sigma_{ij}=(de^\beta/dt)_{k(i}(e^{-\beta})_{j)k}$ and $\sigma^2=\sigma_{ij}\sigma_{ij}$, we must have $\frac{\sigma}{d\Omega/dt}\rightarrow  0$. This last condition means that the anisotropy measured locally through the constant of Hubble tends to zero. It implies that the shear parameter $x\propto \dot\beta_\pm = d\beta_\pm/dt dt/d\Omega\rightarrow 0$. Consequently, isotropisation occurs when $x$ vanishes in $\Omega\rightarrow -\infty$. It is thus a stable state taking place for a diverging value of $t$.\\\\
We deduce that the Universe can become isotropic following three different ways. We name them respectively class 1, 2 and 3. They are defined as follows;
\begin{itemize}
\item Class 1: in the vicinity of the isotropy, all the variables $(x,y,z)$ reach equilibrium with $y\not =0$. It is generally possible to determine the asymptotical forms of the metric functions and potential whatever the forms of $\omega$, $U$ and $\lambda$.
\item Class 2: in the vicinity of the isotropy, all the variables $(x,y,z)$ reach equilibrium with $y=0$ and thus $k^2\rightarrow 1-12z^2$. Until now, we did not succeed in finding the Universe asymptotical state. A numerical example will be shown in the last section.
\item Class 3: in the vicinity of the isotropy, only the variable $x$ reaches equilibrium but not necessarily the others variables. If $y$ and $z$ do not reach equilibrium when $\Omega\rightarrow -\infty$ whereas they are bounded, they necessarily oscillate without damping. Thus their first derivatives oscillate around zero. This phenomenon will arise if $\ell$ or/and $\ell_m$ sufficiently oscillate when $\Omega\rightarrow -\infty$ such that the signs of $\dot y$ or/and $\dot z$ change continuously. Of course in this case, once again it seems difficult to determine the Universe asymptotical state (but we hope not impossible). Class 3 isotropisation has been observed numerically in presence of a complex scalar field in \cite{FayLum03}.
\end{itemize}
In this paper, we will consider the first class which is agreement with the observations. Moreover, itallows a full description of the Universe asymptotic state when it isotropises.
%------------------------------------------------------------------------------------------------------------------------------------------------------------------------------------------%
\subsection{Assumptions for the stability of our results}\label{s20}
In this paper, we will determine the isotropic equilibrium points, some necessary conditions for isotropisation and the asymptotical behaviours of some functions in the neighbourhood of these points. However we will get these behaviours by neglecting two quantities in the vicinity of the equilibrium. The first one is the variation of a function $f$ of the scalar field whose form depends on the asymptotical behaviour of $k$. The second one is the variations of the variables $(y,z,k)$. In other words, we will assume that all these quantities tend sufficiently fast to their equilibrium values. We had already talked about this problem in \cite{a3} and we reproduce below the discussion of this last paper.\\
The first type of assumption is related to $\ell$ and $\ell_m$. Let $f(\ell,\ell_m)$ be a function of the scalar field, that we will define below, and tending to a constant $f_0$ in the vicinity of the isotropic state. We will assume that this function reaches sufficiently quickly its equilibrium value $f_0$, i.e.
\begin{itemize}
\item When $f$ tends to its constant equilibrium value $f_0$ (vanishing or not) such as $f\rightarrow f_0+\delta f$, $\int(f_0+\delta f) d\Omega\rightarrow f_0\Omega+const$.
\end{itemize}
We will check this assumption each time we will use our results. If it is not true, the asymptotical behaviours for the metric functions (and potential) will be different from the laws we will derive below. This problem could be overcame since our results allow to calculate $\phi(\Omega)$ and thus $f(\Omega)$. Hence, it should be easy to generalise them by keeping the $\int f d\Omega$ term instead of considering that it tends to $f_0\Omega+const$. However, they would not be on a closed form.
\\\\
The second type of assumption can not be raised so easily. In the same way, the asymptotical behaviours we will determine will be true only if the variables $(y,z,k)$ tend sufficiently fast to their equilibrium values. It means that we have to make the same kind of assumption for $(y,z,k)$ as for $f$. A solution to solve this problem would be to consider some small perturbations of these variables in the vicinity of the equilibrium. However, until now we did not succeed to get any interesting results, even for an empty flat model.
\\\\
To summarize, the results of this paper concerning the asymptotical behaviours will be valid for a class 1 isotropisation if the functions $f$ defined below and the variables $(y,z,k)$ tend sufficiently fast to their equilibrium values. From a physical point of view, it means that the Universe tends sufficiently fast to its isotropic state. The assumption on $f$ may be easily raised but the ones on $(y,z,k)$ need a more careful examination.
%------------------------------------------------------------------------------------------------------------------------------------------------------------------------------------------%
\subsection{Asymptotic state when $k\not =0$}\label{s22}
In what follows, we look for the isotropic states with $k\not =0$. The case with $k\rightarrow 0$ will be analysed in the section \ref{s222}.\\

\begin{figure}[h]
\begin{center}
\includegraphics[width=\textwidth]{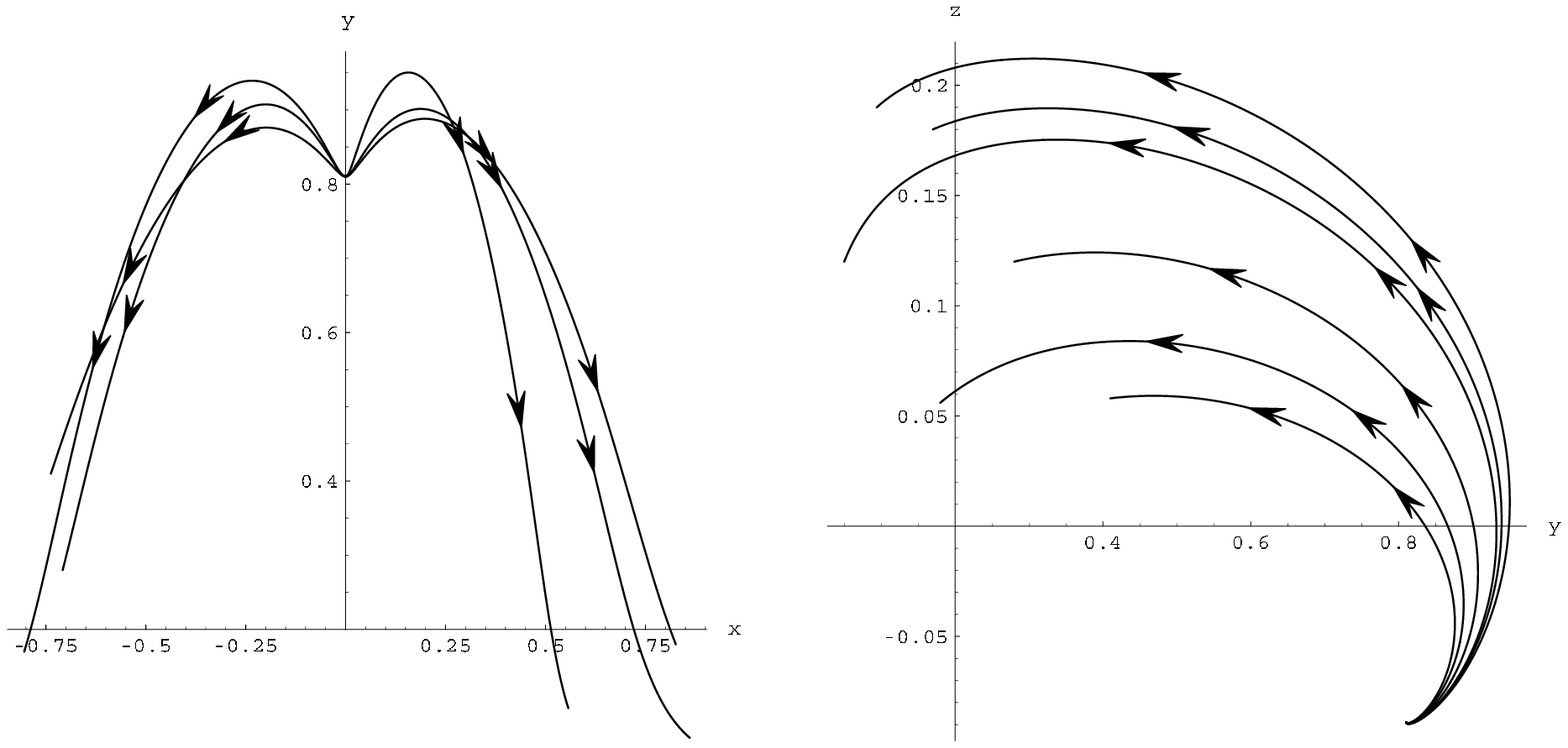}
\caption{\label{fig1}Equilibrium point $E_2$ in $(x,y,z)=(0,0.81,-0.08)$ when $k\not =0$ and $(R,\gamma,\ell,\ell_m)=(1,1,-1.23,1.58)$.}
\end{center}
\end{figure}

\emph{Equilibrium points}\\
First we calculate the equilibrium points of the equations system (\ref{12}-\ref{14}). Then, we introduce them in the constraint (\ref{11}) to find $k$. The equilibrium points write:
%-------------------------EQUATION-------------------------------------%
\begin{eqnarray*}
E_0&=&(0,0,\frac{\ell_m}{3(2-\gamma)})\\
E_1&=&(0,\pm\frac{1}{2\sqrt{3}R(\ell-\ell_m)}\mbox{[}-4\ell^4+8\ell^3\ell_m-4\ell^2(3+\ell_m^2-3\gamma)-\\
&&12\ell\ell_m(\gamma-1)-9\gamma(\gamma-2)\mbox{]}^{1/2},\frac{6\ell+2\ell^3-6\ell_m-2\ell^2\ell_m-3\ell\gamma}{12\ell(\ell-\ell_m)})\\
E_2&=&(0,\pm\frac{1}{2R(\ell-\ell_m)}\left[4\ell_m(\ell_m-\ell)-3\gamma(\gamma-2)\right]^{1/2},\frac{\gamma}{4(\ell-\ell_m)})
\end{eqnarray*}
$E_0$ belongs to the class 2 and is thus discarded. For the two other points, we find for $k$:
$$k^2=\frac{2\ell(\ell-\ell_m)-3\gamma}{2(\ell-\ell_m)^2},$$
$k$ is a real as long as:
%-------------------------EQUATION-------------------------------------%
\begin{equation}\label{c1}
\ell(\ell-\ell_m)>\frac{3}{2}\gamma
\end{equation}
This inequality implies a real value for points $E_1$ which are thus discarded from further considerations. Consequently, the only equilibrium points corresponding to an isotropic class 1 stable state are the $E_2$ ones. A numerical simulation representing one of them is represented on figure \ref{fig1}. 
They are real and bounded if respectively:
%-------------------------EQUATION-------------------------------------%
\begin{equation}\label{c2}
4\ell_m(\ell_m-\ell)>3\gamma(\gamma-2)
\end{equation}%-------------------------EQUATION-------------------------------------%
\begin{equation}\label{c3}
\ell\not\rightarrow \ell_m
\end{equation}
i.e. $U\not\rightarrow \lambda$. The first condition is always satisfied when there is no coupling between the matter and the scalar field($\ell_m=0$). We will show in section \ref{s23} that $\ell$ and $\ell_m$ can not diverge but at the same order and that $k$ tends to a non vanishing constant in the vicinity of the $E_2$ points.\\

\emph{Monotonic functions}\\
The equation (\ref{12}) shows that $x$ is a monotonic function of constant sign. We deduce that the metric functions, whose derivatives with respect to the proper time express as some linear functions of $x$, can only have a single extremum. From the form of the lapse function $N$ and the relation $dt=-N d\Omega$, it comes that $\Omega$ is a monotonic function of the proper time. Its value in $-\infty$ corresponds to late time epoch when the Hamiltonian is initially positive. We will find the same monotonic functions when $k\rightarrow 0$.\\

\emph{Asymptotical behaviours}\\
The function $f$ (see subsection \ref{s20}) is defined in this subsection as $f=\ell(\ell-\ell_m)^{-1}$. Then linearising the equation (\ref{12}) for $x$ in the vicinity of the isotropic equilibrium state, we find:
$$
x\rightarrow x_0 e^{-\frac{3\left[2\ell_m+\ell(\gamma-2)\right]}{2(\ell-\ell_m)}\Omega},
$$
with $x_0$ an integration constant\footnote{This result is always valid if $\ell$ and $\ell_m$ diverge at the same order.}. $x$ vanishes when $\Omega\rightarrow -\infty$ if the reality conditions for $k$ and $E_2$ are respected. From the lapse function $N$ and the relation $dt=-Nd\Omega$, we find that $e^{-\Omega}$ tends to the increasing function of the proper time:
%-------------------------EQUATION-------------------------------------%
\begin{equation}\label{r1}
e^{-\Omega}\rightarrow t^{\frac{2(\ell-\ell_m)}{3\ell\gamma}}
\end{equation}
when $\frac{3\ell\gamma}{2(\ell-\ell_m)}$ tends to a non vanishing constant. This is always true since, as it will be shown in the next section, $\ell$ and $\ell_m$ can not diverge but at the same order with $\ell\not\rightarrow \ell_m$. Moreover, the reality condition (\ref{c1}) for $k$ shows that $\ell$ does not vanish. Since in the next section we will also prove that $y$ can not vanish, from $y$ definition we derive that the potential asymptotically disappears as $U\rightarrow t^{-2}$.\\
To check the necessary conditions for isotropy (\ref{c1}-\ref{c3}) when $\omega(\phi)$ and $U(\phi)$ are specified, we need to know the asymptotical behaviour of $\phi$. For this, we write (\ref{15}) in the neighbourhood of the equilibrium. Then, we get that asymptotically $\phi$ behaves as the asymptotical solution of the differential equation:
%-------------------------EQUATION-------------------------------------%
\begin{equation}\label{dotphi}
\dot\phi=3\gamma(\frac{U_\phi}{U}-\frac{\lambda_\phi}{\lambda})^{-1}
\end{equation}
when $\Omega\rightarrow -\infty$. This asymptotical equation is a good illustration of the assumptions on $(y,z,k)$ described in subsection \ref{s20}: to determine (\ref{dotphi}), we have replaced $z$ in (\ref{15}) by its equilibrium value $\frac{\gamma}{4(\ell-\ell_m)}$, neglecting any small variation $\delta z$ when $z$ approaches equilibrium. However, if it tends to $\frac{\gamma}{4(\ell-\ell_m)}$ slower than $\Omega^{-1}$, then $\delta z$ should be taken into account in (\ref{dotphi}) instead of being neglected.
%------------------------------------------------------------------------------------------------------------------------------------------------------------------------------------------%
\subsection{Some important results for the $E_2$ points}\label{s23}
Integrating (\ref{dotphi}) near equilibrium leads to 
$$
U\rightarrow U_0\lambda V^{-\gamma}
$$
$U_0$ being an integration constant. We then deduce from the expression (\ref{form2}) that $k$ tends to a non vanishing constant as long as $y\not =0$ what is always true. Indeed, the only ways for $y$ to vanish are if $\ell>>\ell_m$ or $4\ell_m(\ell_m-\ell)\rightarrow 3\gamma(\gamma-2)$. In the first case, the constraint or equivalently (\ref{r1}) shows that $k\rightarrow 1$. But if we consider the form (\ref{form2}) for $k$ and the limit $U\rightarrow U_0\lambda V^{-\gamma}$, it comes that if $y\rightarrow 0\Rightarrow k\rightarrow 0\not=1$. For the same reason, $y$ can not tend to $0$ when $4\ell_m(\ell_m-\ell)\rightarrow 3\gamma(\gamma-2)$. It follows that $y$ is never vanishing in the vicinity of $E_2$. We get a similar result if we consider the divergence of $\ell_m$ when $\ell_m>>\ell$. In this case, $y$ tends to a non vanishing constant and $k$ to $0$, which is in disagreement with the form (\ref{form2}) of $k$ and the limit $U\rightarrow U_0\lambda V^{-\gamma}$. On the other hand, if $\ell$ and $\ell_m$ diverge at the same order without converging one to the other, $y$ and $k$ tend to some non vanishing constants and the constraint is respected. Since $\lambda\propto UV^{\gamma}$ and $y$ is not vanishing, we deduce from (\ref{var2}) that
$$\lambda\rightarrow e^{\frac{3\gamma\ell_m\Omega}{\ell-\ell_m}}$$
and from (\ref{r1}) that 
$$\lambda\rightarrow t^{-2\frac{\ell_m}{\ell}}$$
Thus from the reality condition for $k^2$, it comes $\lambda>t^{-2(1-\frac{3}{2}\frac{\gamma}{\ell^2})}$. In the same time, the energy densities for the perfect fluid and scalar field write respectively $\rho=\lambda V^{-\gamma}\rightarrow U$ and $\rho_\phi=\frac{9}{2}\gamma^2(\ell^{-1}-\ell_m^{-1})^{-2}t^{-2}+\frac{1}{2}U$. When $\ell$ and $\ell_m$ tend to some constants, since $U\rightarrow t^{-2}$, $p_\phi\propto\rho_\phi\propto\rho_m$: the scalar field and the perfect fluid energy densities behave in the same way. If both $\ell$ and $\ell_m$ diverge at the same order, the kinetic term in $\rho_\phi$ is larger than $t^{-2}$, $\rho_\phi>>\rho_m$ and the energy density of the scalar field dominates.
%------------------------------------------------------------------------------------------------------------------------------------------------------------------------------------------%
\subsection{Equilibrium point when $k\rightarrow 0$}\label{s222}
This section is divided in three parts depending on $\ell_m=0$ strictly, $\ell_m k^2\rightarrow 0$ or $\ell_m k^2\not\rightarrow 0$.\\
\\
\underline{$\ell_m=0$}\\
In this first part, we recall and complete the results got in \cite{Fay01A} when no coupling exists between the scalar field and the perfect fluid.\\
When there is no perfet fluid, $k=0$ strictly and the reality condition for the equilibrium points writes $\ell^2<3$. Near isotropy, the metric functions tend to $t^{\ell^{-2}}$ when $\ell$ tends to a non vanishing constant or to an exponential when $\ell$ vanishes. When $k\not\rightarrow 0$, the reality condition for the equilibrium points is $\ell^2>3/2\gamma$ and the metric functions tend to $t^{\frac{2}{3\gamma}}$. When $k\rightarrow 0$, we recover the same values for the equilibrium points as when no perfect fluid is present\cite{Fay01} but now, $k\rightarrow 0$ implies $\ell^2<3/2\gamma$. The asymptotical behaviour of the metric functions is the same as without a perfect fluid. We had not noticed this last inequality in \cite{Fay01A} nor that $U\rightarrow V^{-\gamma}$ when $k\rightarrow const\not =0$.\\
Once again, these results rest on the assumptions of subsection \ref{s20} with now $f=\ell^2$ when $k=0$ or $k\rightarrow 0$. When they are not true, meaning that the Universe does not reach its isotropic equilibrium state sufficiently quickly, the asymptotical behaviours of $U$ and $e^{-\Omega}$ are generally different. As instance when $k=0$ strictly and if $\ell^2$ vanishes as $n\Omega^{-1}$ with $n<0$, the integral of $f$ does not tend to a constant. Then we can show that the potential will diverge as $(-\Omega)^{-2n}$ and the metric functions will tend to $exp\left[\frac{n+1}{12\pi R_0^3 x_0}t^{1/(n+1)}\right]$ with $n\in\left]-1,0\right[$ such as the Universe be expanding. This solution is different from the classical solutions found when we neglect the variation of $\ell$ near equilibrium. It shows that the assumptions on $f$ that we will also use in the next sections, have to be checked each time we apply our results to a specific scalar-tensor theory.
\\
\\
\underline{$\ell_m k^2\rightarrow 0$}\\
If $k\rightarrow 0$ with $\ell_m k^2\rightarrow 0$, again we recover the same equilibrium points and behaviour for $x$ as in the absence of a perfect fluid. That is $x\rightarrow x_0e^{(3-\ell^2)\Omega}$ with $\ell^2<3$ such that $x\rightarrow 0$ in $\Omega\rightarrow -\infty$ and the equilibrium points are real. Consequently, using the form (\ref{form1}) for $k^2$, it comes $k^2\rightarrow \lambda e^{2(3/2\gamma-\ell^2)\Omega}$ when $\Omega\rightarrow -\infty$. When $\ell^2$ tends to a non vanishing (vanishing) constant smaller than $3$, $e^{-\Omega}\rightarrow t^{\ell^{-2}}$ (respectively $e^{-\Omega}\rightarrow e^{(12\pi R_0^3 x_0)^{-1} t}$). Hence, $k\rightarrow 0$ if
%-------------------------EQUATION-------------------------------------%
\begin{equation}\label{lim10}
\lambda e^{2(3/2\gamma-\ell^2)\Omega}\rightarrow 0
\end{equation}
and thus $\lambda<t^{-2(1-\frac{3}{2}\frac{\gamma}{\ell^2})}$ (respectively $\lambda<e^{3\gamma(12\pi R_0^3x_0)^{-1} t}$). In the same way, $\ell_m k^2\rightarrow 0$ if:
%----------%
$$\ell_m \lambda e^{2(3/2\gamma-\ell^2)\Omega}\rightarrow 0$$
%----------%
Contrary to the case $\ell_m=0$, the condition $k\rightarrow 0$ does not automatically restrict the set of $\ell$ allowing isotropisation: it is the form of $\lambda$ which will lay down the law. Since we consider a class 1 isotropisation such as $y\not =0$ and $k\rightarrow 0$, we have $\lambda V^{-\gamma}<<U$ and thus $\rho_\phi-p_\phi>>\rho_m$: the scalar field energy density dominates the Universe.\\
The asymptotical behaviour of the scalar field when $\Omega\rightarrow -\infty$ is given by\cite{Fay01}:\\
%-------------------------EQUATION-------------------------------------%
\begin{equation}
\dot{\phi}=2\frac{\phi^2U_\phi}{U(3+2\omega)}
\end{equation}
\underline{$\ell_m k^2\not\rightarrow 0$}\\
Since $k\rightarrow 0$ whereas $\ell_m k^2\not\rightarrow 0$, it means that $\ell_m$ have to diverge. The equilibrium points when $\ell_m k^2\not\rightarrow 0$ write:
$$
E_3=(0,\pm R^{-1},0)
$$
They are such as $k^2=-\ell\ell_m^{-1}$. The dynamics approach these points in the same way as on the figure \ref{fig1}. $k$ will vanish and will be real if respectively $\ell<<\ell_m$ and $\ell\ell_m^{-1}<0$. Moreover, $\ell_m k^2\not\rightarrow 0$ if $\ell$ tends to a non vanishing constant or diverges with $z\ell$ bounded. Mathematically, the $E_3$ point could be the asymptotical limit of $E_2$ when $\ell_m$ diverges and $\ell<<\ell_m$. However, this divergence is forbidden by the constraint.\\
Near $E_3$, we find that $x\rightarrow e^{3\Omega}$, showing that the Universe tends to a De Sitter model, i.e. $e^{-\Omega}\rightarrow e^{(12\pi R_0^3 x_0)t}$, and the potential to the constant $(R x_0)^{-2}$. As previously, we have then $k\rightarrow 0$ if
$$\lambda e^{3\gamma\Omega}\rightarrow 0$$
i.e. $\lambda<e^{3\gamma(12\pi R_0^3x_0)^{-1} t}$. In the same way, $\ell_m k^2$ does not vanish if:
$$\ell_m \lambda e^{3\gamma\Omega}\not\rightarrow 0$$
Again, $y$ being different from $0$ and considering the form (\ref{form2}) for $k^2$, we have $\rho_m=\lambda V^{-\gamma}<<U$ such as $k\rightarrow 0$ and thus $\rho_m<<\rho_\phi-p_\phi$: the Universe is scalar field dominated. From (\ref{form1}) and the limit of $k$ near equilibrium, we determine the scalar field asymptotical behaviour:
%-------------------------EQUATION-------------------------------------%
\begin{equation}
\delta\frac{1}{\lambda_\phi}\frac{U_\phi}{U}=e^{3\gamma\Omega}
\end{equation}
%------------------------------------------------------------------------------------------------------------------------------------------------------------------------------------------%
\section{Discussion}\label{s3}
The discussion is divided in three parts. In the first one we summarize our results and in the second one we consider some applications for Brans-Dicke and low energy string theories. We conclude in the third one.
%------------------------------------------------------------------------------------------------------------------------------------------------------------------------------------------%
\subsection{Summary}\label{s31}
We have studied the necessary conditions which may lead the Universe to a class 1 isotropisation. It depends if $k$ does not vanish, vanishes with $\ell_m k^2\rightarrow 0$ or with $\ell_m k^2\not\rightarrow 0$. We assumed that $3+2\omega$ and $U$ are some positive functions of the scalar field and that the isotropic state was reached sufficiently fastly. Below we summarize our results.\\
\\
\emph{\underline{Case 1: $k\not \rightarrow 0$}\\
We define the quantities $\ell=\phi U_\phi U^{-1}(3+2\omega)^{-1/2}$ and $\ell_m=\phi \lambda_\phi \lambda^{-1}(3+2\omega)^{-1/2}$. Let $p_\phi$, $\rho_\phi$ and $\rho_m$ be respectively the pressure and density of the scalar field, the density of the perfect fluid. Some necessary conditions for Bianchi type $I$ isotropisation in presence of a massive scalar field minimally coupled to the curvature but not minimally coupled to the perfect fluid are:
\begin{itemize}
\item $\ell\not\rightarrow \ell_m$ (equilibrium points are bounded)
\item $4\ell_m(\ell_m-\ell)>3(\gamma-2)\gamma$ (reality condition)
\item $\ell(\ell-\ell_m)>\frac{3}{2}\gamma$ (reality condition)
\item $\ell$ and $\ell_m$ are bounded or diverge in the same way (the constraint is respected)
\end{itemize}
When isotropy is approached, the metric functions behave as $t^{\frac{2(\ell-\ell_m)}{3\ell\gamma}}$, $\lambda\rightarrow t^{-2\frac{\ell_m}{\ell}}$ whereas the potential decreases as $t^{-2}$. When $\ell$ and $\ell_m$ do not diverge, there is an equilibrium between the scalar field and the perfect fluid: $\rho_\phi\propto p_\phi\propto\rho_m$. When both diverge, $\rho_\phi-p_\phi>>\rho_m$ and the Universe is scalar field dominated. Asymptotically, the scalar field checks the relation $U\rightarrow U_0\lambda e^{3\gamma\Omega}$.
}\\
\\
This last relation determines the asymptotical form of $\phi$ and thus these of $\ell$ and $\ell_m$. Note that in the case $\ell_m=0$\cite{Fay01A}, the metric functions asymptotical behaviour does not depend on $\phi$ and is always $t^{\frac{2}{3\gamma}}$. It thus prevents any late time acceleration. Hence, it is the existence of a coupling between the scalar field and the perfect fluid which allows the appearance of an accelerated expansion when $k\rightarrow const \not =0$. Then, since when $\ell$ and $\ell_m$ are bounded we have $\rho_\phi\propto \rho_m$, it follows that $\Omega_\phi\propto \Omega_m$ and the coincidence problem could be solved.
\\\\
\emph{\underline{Case 2: $k\rightarrow 0$ and $\ell_m k^2\rightarrow 0$}\\
We define the quantities $\ell=\phi U_\phi U^{-1}(3+2\omega)^{-1/2}$ and $\ell_m=\phi \lambda_\phi \lambda^{-1}(3+2\omega)^{-1/2}$. Some necessary conditions for Bianchi type $I$ isotropisation in presence of a massive scalar field minimally coupled to the curvature but not minimally coupled to the perfect fluid are:
\begin{itemize}
\item $\ell^2<3$ (reality condition)
\item $\lambda e^{2(3/2\gamma-\ell^2)\Omega}\rightarrow 0$ (Condition for $k\rightarrow 0$)
\item $\ell_m \lambda e^{(3\gamma-2\ell^2)\Omega}\rightarrow 0$ (Condition for $\ell_m k^2\rightarrow 0$)
\end{itemize}
If $\ell^2$ tends to a non vanishing constant, the metric functions tend to $t^{\ell^{-2}}$ and the potential vanishes as $t^{-2}$. If $\ell^2$ vanishes, the Universe tends to a De Sitter model and the potential to a constant. In any cases $\rho_\phi-p_\phi>>\rho_m$ and the Universe is scalar field dominated. The asymptotical behaviour for the scalar field is this of the asymptotical  solution of $\dot{\phi}=2\frac{\phi^2U_\phi}{U(3+2\omega)}$.
}\\
\\
These results include the ones got when there is no perfect fluid\cite{Fay01}. For sake of clarity, we chose to express the above (as well as below) limits for $k\rightarrow 0$ and $\ell_m k\rightarrow 0$ as some functions of $e^{-\Omega}$ and $\phi$. However, these two last quantities are asymptotically defined with respect to the proper time $t$ by the behaviours of the metric functions and potential.\\
\\
\emph{\underline{Case 3: $k\rightarrow 0$ and $\ell_m k^2\not\rightarrow 0$}\\
We define the quantities $\ell=\phi U_\phi U^{-1}(3+2\omega)^{-1/2}$ and $\ell_m=\phi \lambda_\phi \lambda^{-1}(3+2\omega)^{-1/2}$. Some necessary conditions for Bianchi type $I$ isotropisation in presence of a massive scalar field minimally coupled to the curvature but not minimally coupled to the perfect fluid are:
\begin{itemize}
\item $\ell_m$ diverges and $\ell\rightarrow const \not = 0$ or diverges such that $z\ell\rightarrow 0$ (condition for $\ell_mk^2\rightarrow 0$)
\item $\ell<<\ell_m$ or $\lambda e^{3\gamma\Omega}\rightarrow 0$ (condition for $k\rightarrow 0$)
\item $\ell\ell_m^{-1}<0$ (reality condition)
\end{itemize}
The Universe tends to a De Sitter model and the potential to a constant. Since $\rho_\phi-p_\phi>>\rho_m$, the scalar field asymptotically dominates the Universe and checks the equation $\delta\frac{1}{\lambda_\phi}\frac{U_\phi}{U}=e^{3\gamma\Omega}$.
}\\
\\
The cases with $k\not\rightarrow 0$ and $k\rightarrow 0$ are strictly separated by the asymptotical behaviour of $\lambda$ since the first one implies $\lambda>t^{-2(1-\frac{3}{2}\frac{\gamma}{\ell^2})}$ and the second one $\lambda<t^{-2(1-\frac{3}{2}\frac{\gamma}{\ell^2})}$ (or $\lambda<e^{3\gamma(12\pi R_0^3 x_0)^{-1}t}$ when $\ell\rightarrow 0$). The two cases with $k\rightarrow 0$ are distinguished by the fact that for the first one, the Universe tends to a De Sitter model when $\ell\rightarrow 0$ and the second one when $\ell\not =0$.
%------------------------------------------------------------------------------------------------------------------------------------------------------------------------------------------%
\subsection{Applications}\label{s32}
\begin{figure}[h]
\includegraphics[width=\textwidth]{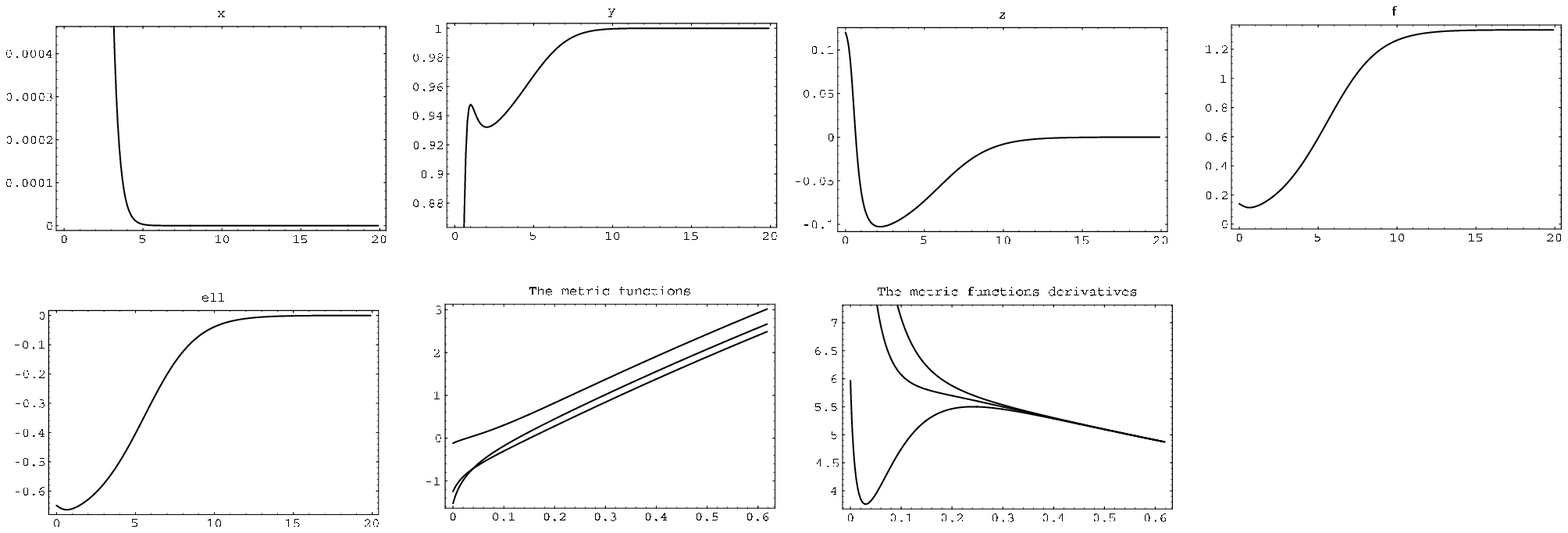}
\caption{\scriptsize{\label{fig2}These figures represent the approach for a class 1 isotropisation when $\omega_0=2.3$, $n=1.5$ and $m=1.1$. As expected, $x$ tends to $0$, $\phi$ to the constant $2/n=1.33$ and $\ell$ (here named ell) to $0$. The convergence of $\phi$ to a constant is in accordance with the fact that $U$ also tends to a constant and the Universe to a De Sitter model. In the Brans-Dicke frame, the derivatives of the metric functions $\alpha$, $\beta$ and $\gamma$ tend to the same behaviour: isotropisation occurs.}}
\end{figure}

\begin{figure}[h]
\includegraphics[width=\textwidth]{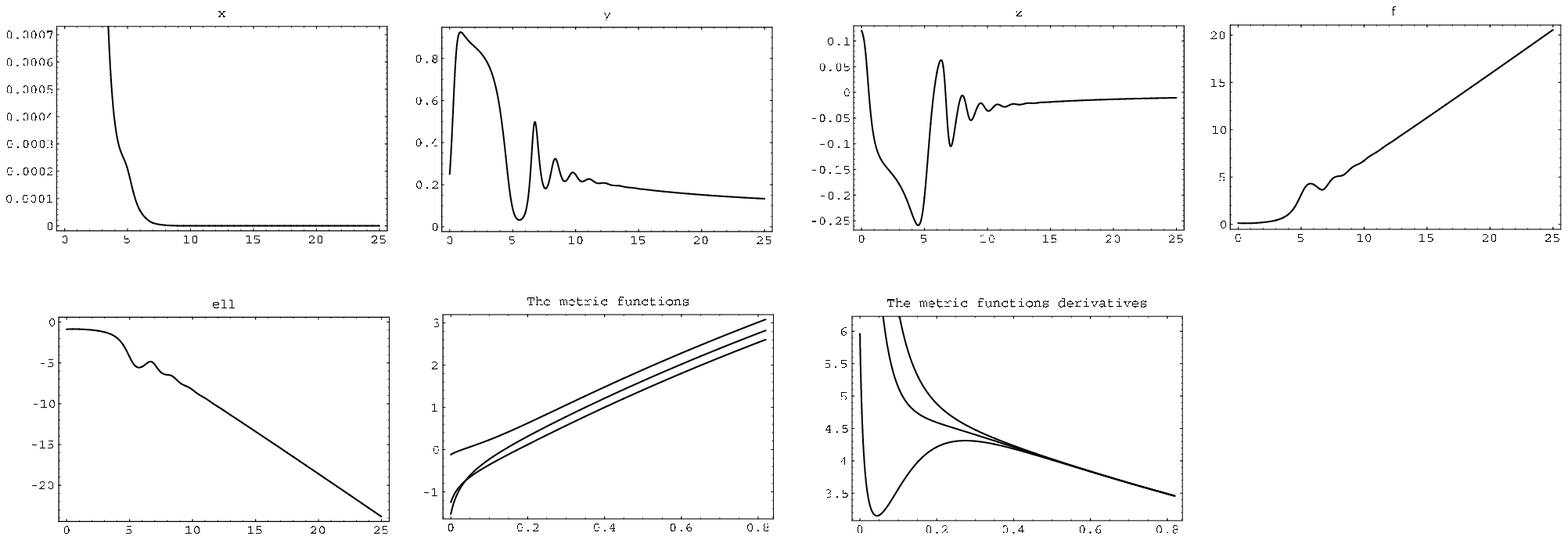}
\caption{\scriptsize{\label{fig3}These figures represent the approach for a class 2 isotropisation when $\omega_0=2.3$, $n=-3.1$ and $m=1.1$. $x$ always tends to $0$ but also $y$. $\phi$ and thus $\ell$ diverge. Note that $\phi$, $y$, $z$ and $\ell$ undergo damped oscillations. In the Brans-Dicke frame, the derivatives of the metric functions tend to the same behaviour showing isotropisation.}}
\end{figure}

\begin{figure}[h]
\includegraphics[width=\textwidth]{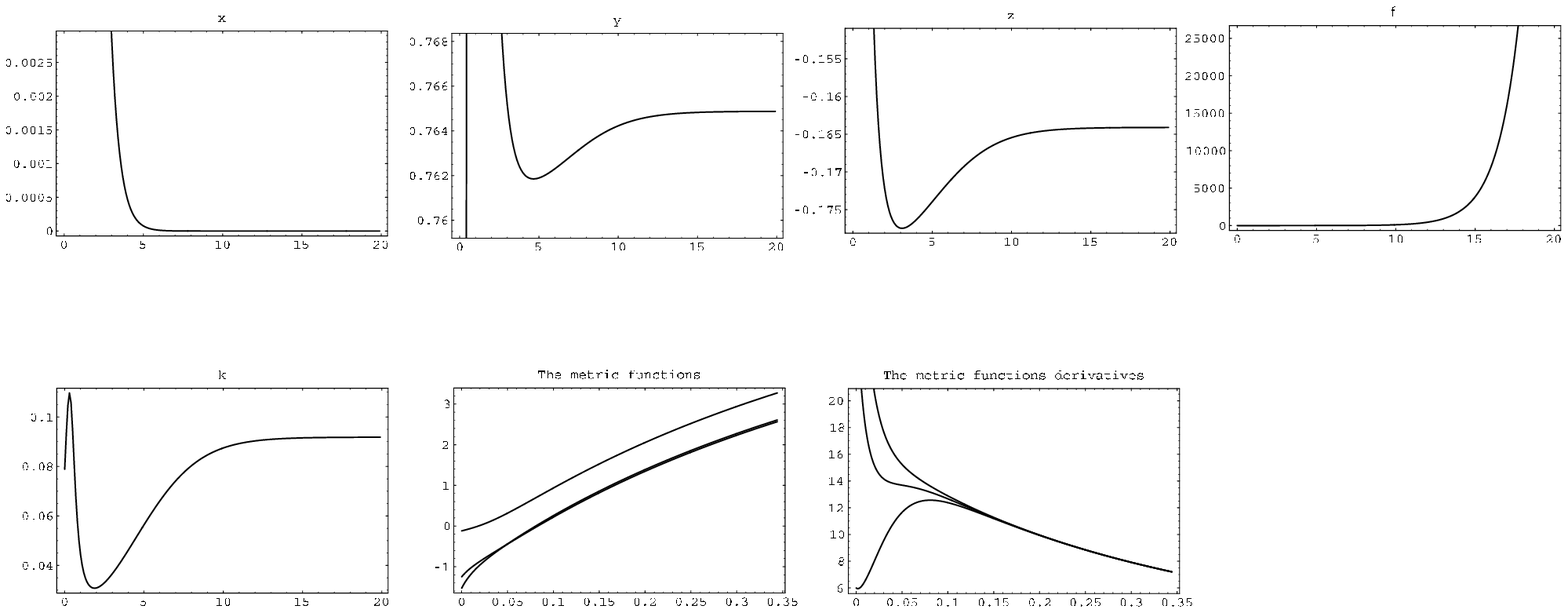}
\caption{\scriptsize{\label{fig4}These figures represent the approach for a class 1 isotropisation when $\omega_0=2.3$, $n=-3.1$ and $m=1.1$. Anew $x$ tends to vanish, $y$ and $k$ to some non vanishing constants showing that $U\propto \lambda e^{3\gamma\Omega}$. $\phi$ diverges since $m-n>0$. In the Brans-Dicke frame, the Universe isotropises.}}
\end{figure}

\begin{figure}[h]
\includegraphics[width=\textwidth]{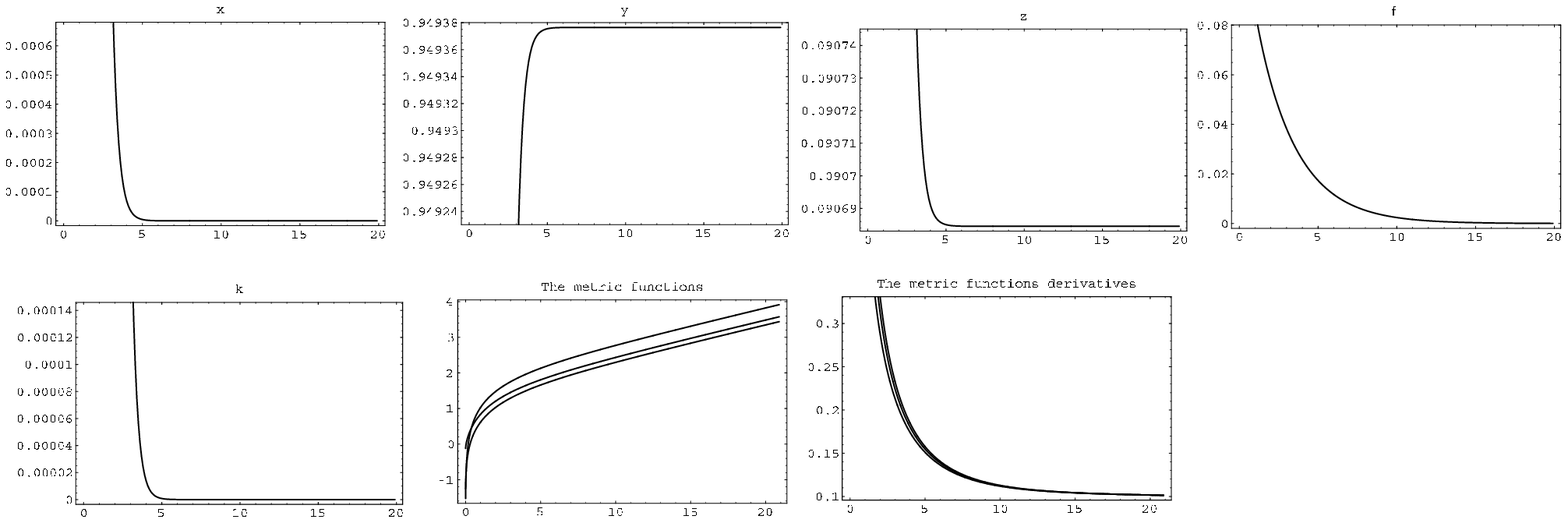}
\caption{\scriptsize{\label{fig5}These figures represent the approach for a class 1 isotropisation when $\omega_0=2.3$, $n=1.5$ and $m=1.1$. Here, k tends to vanish and the scalar field energy density dominates the Universe.}}
\end{figure}

\begin{figure}[h]
\includegraphics[width=\textwidth]{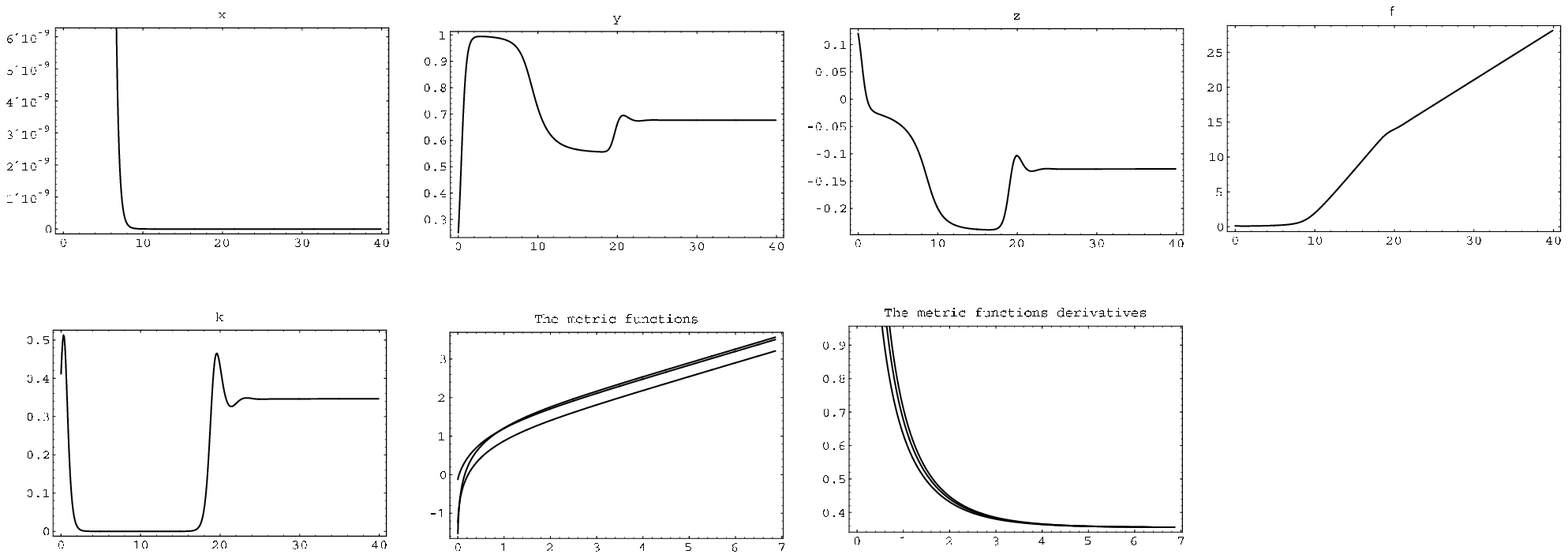}
\caption{\scriptsize{\label{fig6}These figures represent the approach for a class 1 isotropisation when $\omega_0=2.3$, $n=-3.1$ and $m=1.1$. k tends to a constant showing the equilibrium between the scalar field and the perfect fluid. Remark the existence before this equilibrium of a period during which the density of the scalar field dominated the one of the perfect fluid.}}
\end{figure}

In what follows, we apply our results to some Brans-Dicke and low energy string actions. For that, we use a conformal transformation of the metric described in the appendice. It casts the minimally coupled scalar-tensor theory (\ref{action}) in the Einstein frame where our results take place, into a non minimally coupled scalar-tensor theory (\ref{lbd}) in the Brans-Dicke frame. Obviously, when isotropy arises in the Einstein frame, it also occurs in the Brans-Dicke frame. Thus the necessary conditions for isotropy are the same in both frames. However, the metric functions generally behave differently.\\
We will illustrate each of the four applications below with some figures representing the behaviours of $x$, $y$, $z$, $k$, $\phi$ and $\ell$ in the Einstein frame and in the $\Omega$ time. The numerical integrations will be done using the initial conditions $\phi_0=0.14$, $y_0=0.25$, $z_0=0.12$. $x_0$ is calculated with the constraint (\ref{11}) where we put $p_+^2+p_-^2=p^2=1$, $R=1$ and $\delta=1$ (the constant in the definition of $k$). The behaviours of the metric functions $\alpha$, $\beta$ and $\gamma$ in the Brans-Dicke frame, and their derivatives, will be also represented but in the proper time $t$ with initial conditions $\alpha_0=-1.53$, $\beta_0=-1.25$, $\gamma_0=0.12$, $d\alpha_0/d\tau_0=2.48$, $d\beta_0/d\tau_0=1.55$ and $d\gamma/d\tau_0=0.33$. Here, the $\tau$ time is defined as $dt=V d\tau$. To get the figures, we numerically integrated the field equations of the Lagrangian formalism using a 5 order Runge-Kutta method\footnote{We used java oriented object programing to perform the calculations.}. $d\phi_0/d\tau_0$ is calculated using the constraint equation of this formalism. Each time, a dust fluid and a null initial time were asumed.
\\\\
\textbf{\emph{Brans-Dicke theory with an exponential potential}}\\
Consider the class of theories defined by (\ref{action}) such that:
%-------------------------EQUATION-------------------------------------%
\begin{eqnarray*}
\omega&=&\omega_0\\
U&=&\phi^{-2}e^{n\phi}\\
\lambda&=&\phi^m
\end{eqnarray*}
Using the conformal transformation, it can be cast into the non minimally coupled scalar field theory (\ref{lbd}) defined by:
%-------------------------EQUATION-------------------------------------%
\begin{eqnarray*}
G&=&\phi^{\frac{m}{3(4-3\gamma)}}\\
\omega&=&\left[\frac{3}{2}(1-\frac{m^2}{9(4-3\gamma)^2})+\omega_0)\right]\phi^{\frac{-m}{3(4-3\gamma)}-1}\\
U&=&\phi^{-2(1+\frac{m}{3(4-3\gamma)})}e^{n\phi}
\end{eqnarray*}
The Brans-Dicke theory with an exponential potential is recovered for $m=3(3\gamma-4)$.\\
The quantities $\ell$ and $\ell_m$ are defined by:
$$\ell=\frac{n\phi-2}{\sqrt{3+2\omega_0}}$$
$$\ell_m=\frac{m}{\sqrt{3+2\omega_0}}$$
$\ell_m$ can not diverge and consequently the \underline{case 3} never happens. Moreover $3+2\omega_0$ have to be positive. For the \underline{case 1}, near the isotropic equilibrium state, we have for the scalar field:
$$e^{n\phi}\phi^{-(2+m)}\rightarrow U_0e^{3\gamma\Omega}$$
Since $\ell$ is bounded, $\phi$ can not diverge and should asymptotically vanish. It implies that $m<-2$ and finally $\phi\rightarrow e^{\frac{3\gamma}{-(2+m)}\Omega}$. The second reality condition gives:
$$\frac{4(2+m)-3\gamma(3+2\omega_0)}{2(3+2\omega_0)}>0$$
But since $m<-2$, $\gamma>0$ and $3+2\omega_0>0$, it can not be satisfied. Consequently, a class 1 isotropisation does not arise.\\
\\
Let us consider the \underline{case 2}. Integrating the differential equation for $\phi$, we get:
$$\phi\rightarrow \frac{2}{n-\phi_0 e^{\frac{4\Omega}{3+2\omega_0}}}$$
Then, when $\Omega\rightarrow -\infty$, $\phi\rightarrow 2n^{-1}$, $\ell\rightarrow 0$ and $\lambda$ tends to the constant $(2n^{-1})^m$, implying $n>0$. If the Universe isotropises, it will tend to a De Sitter one. Remark that $\phi$, and thus $n$, have to be positive such that $\lambda$ be a real function.\\
\\
Using the conformal transformation, when isotropisation occurs in the Brans-Dicke frame where $\phi$ is non minimally coupled to the curvature and since $\lambda$ tends to a constant, the metric also tends to a De Sitter form (see figure \ref{fig2}).\\
A class 2 isotropisation is also possible when $n<0$ and is plotted on figure \ref{fig3}. As above noted, such a range for $n$ is impossible for a class 1 isotropisation since $\lambda$ would be a complex function. It is the only example of class 2 isotropisation we have found until now.
\\
\\
\textbf{\emph{Brans-Dicke theory with a power potential}}\\
Consider the class of theories defined by the Lagrangian (\ref{action}) such that: 
%-------------------------EQUATION-------------------------------------%
\begin{eqnarray*}
\omega&=&\omega_0\\
U&=&\phi^{n}\\
\lambda&=&\phi^m
\end{eqnarray*}
If we apply again the conformal transformation, we obtain the non minimally coupled scalar tensor theory defined by:
%-------------------------EQUATION-------------------------------------%
\begin{eqnarray*}
G&=&\phi^{\frac{m}{3(4-3\gamma)}}\\
\omega&=&\left[\frac{3}{2}(1-\frac{m^2}{9(4-3\gamma)^2})+\omega_0)\right]\phi^{\frac{-m}{3(4-3\gamma)}-1}\\
U&=&\phi^{n-\frac{2m}{3(4-3\gamma)}}
\end{eqnarray*}
The Brans-Dicke theory with a power law potential is recovered for $m=3(3\gamma-4)$.\\
We calculate that:
$$\ell=\frac{n}{\sqrt{3+2\omega_0}}$$
$$\ell_m=\frac{m}{\sqrt{3+2\omega_0}}$$
with $3+2\omega_0>0$. Anew $\ell_m$ can not diverge and the \underline{case 3} is excluded. For the \underline{case 1}, it is necessary that $n\not = m$ such that $\ell\not\rightarrow \ell_m$. Asymptotically the scalar field behaves as:
$$\phi\rightarrow \phi_0e^{-\frac{3\gamma}{m-n}\Omega}$$
Consequently, in $\Omega\rightarrow -\infty$, $\phi\rightarrow 0$($\phi$ diverges) if $m-n<0$ (respectively $m-n>0$). The reality conditions write:
$$4m(m-n)+3\gamma(2-\gamma)(3+2\omega_0)>0$$
$$2n(n-m)-3\gamma(3+2\omega_0)>0$$
The second one will be respected if n>0(n<0) when $\phi\rightarrow 0$(respectively $\phi$ diverges). We find then that if an isotropic state is reached, the metric functions tend to $t^{\frac{2(n-m)}{3n\gamma}}$ and $\lambda$ to $t^{-\frac{2m}{n}}$.\\
\\
Using the conformal transformation, we deduce for the non minimally coupled theory that the metric functions will tend to:
$$t^{\frac{m(8-5\gamma)+2n(3\gamma-4)}{\gamma\left[m+3n(3\gamma-4)\right]}}$$
All these behaviours are illustrated on figure \ref{fig4}.
\\
\\
For the \underline{case 2}, we get for $\phi$:
$$\phi\rightarrow e^{\frac{2n}{3+2\omega_0}\Omega}$$
Hence $k$ will vanish when $\Omega\rightarrow -\infty$ if $2n(m-n)+3\gamma(3+2\omega_0)>0$. The reality condition for the equilibrium points will be respected if $n^2(3+2\omega_0)^{-1}<3$. The metric functions then tend to $t^{(3+2\omega_0)n^{-2}}$ when $n\not =0$ or to a De Sitter model when $n=0$.\\\\
In the Brans-Dicke frame where the scalar field is non minimally coupled to the curvature, the metric function will tend to:
$$t^{\frac{mn+3(3\gamma-4)(3+2\omega_0)}{n\left[m+3n(3\gamma-4)\right]}}$$
when $n\not =0$. If $n=0$, the behaviour of the metric functions is the same as in the Einstein frame and the Universe tends to a De Sitter model. This case is illustrated on figure \ref{fig5}
\\
\\
\textbf{\emph{Low energy string theory with an exponential potential}}\\
We consider the theory defined by (\ref{action}) and such that: 
%-------------------------EQUATION-------------------------------------%
\begin{eqnarray*}
\omega&=&\omega_0\phi^2+\omega_1\\
U&=&e^{n\phi}\\
\lambda&=&e^{m\phi}
\end{eqnarray*}
Applying the conformal transformation, we define the following non minimally coupled scalar tensor theory:
%-------------------------EQUATION-------------------------------------%
\begin{eqnarray*}
G&=&e^{\frac{m}{3(4-3\gamma)}\phi}\\
\omega&=&\left[\frac{\frac{3}{2}+\omega_0\phi^2+\omega_1}{\phi^{2}}-\frac{3m^2}{18(4-3\gamma)^2}\right]\phi e^{\frac{-m}{3(4-3\gamma)}\phi}\\
U&=&e^{(n-\frac{2m}{3(4-3\gamma)})\phi}
\end{eqnarray*}
The low energy string theory with an exponential potential is then recovered when $m=3(4-3\gamma)$, $\omega_0=5/2$ and $\omega_1=-3/2$.\\
We calculate $\ell$ and $\ell_m$ and we obtain:
$$\ell=\frac{n\phi}{\sqrt{3+2\phi^2\omega_0+2\omega_1}}$$
$$\ell_m=\frac{m\phi}{\sqrt{3+2\phi^2\omega_0+2\omega_1}}$$
These expressions show that we will never have $\ell<<\ell_m$ and thus the \underline{case 3} never occurs. For the \underline{case 1}, it is necessary that $m\not =n$. Moreover, we find for the scalar field:
$$\phi\rightarrow \phi_0+\frac{3\gamma\Omega}{n-m}$$
Hence, $\phi$ diverges and $\ell$ and $\ell_m$ tend to some constants. They will be real if $\omega_0>0$. The reality conditions write:
$$2m(m-n)+3(2-\gamma)>0$$
$$n(n-m)-3\gamma\omega_0>0$$
$\omega_0$ being positive, the second condition needs $n(n-m)>0$ and thus $n\not =0$. Consequently, when isotropisation arises, the metric functions and $\lambda$ respectively tend to $t^{2\frac{n-m}{3n\gamma}}$ and $t^{-2\frac{m}{n}}$.\\
\\
We deduce that in the Brans-Dicke frame, when isotropisation arises, the metric functions will tend to:
$$t^{\frac{m(8-5\gamma)+2n(3\gamma-4)}{\gamma\left[m+3n(3\gamma-4)\right]}}$$
This case is represented on the figure \ref{fig6}.
\\
\\
Concerning the \underline{case 2}, the scalar field asymptotically behaves as:
$$\phi\rightarrow \frac{2n(\Omega-\phi_0)\pm\sqrt{8\omega_0(3+2\omega_1)+4n^2(\phi_0-\Omega)^2}}{4\omega_0}$$
Consequently, depending on the sign of the square root, we have two branches such that $\phi\rightarrow 0$ or $\phi\rightarrow n\omega_0^{-1}\Omega$. For the first one, $\ell\rightarrow 0$ and the Universe tends to a De Sitter model. The limit allowing $k\rightarrow 0$ is always respected. For the second one, $\ell\rightarrow n(2\omega_0)^{-1/2}$ and thus, isotropisation needs $\omega_0>0$ and $n^2(2\omega_0)^{-1}<3$. If $n\not =0$, the metric functions tend to $t^{\frac{2\omega_0}{n^2}}$ and the limit allowing $k\rightarrow 0$ is satisfied if $\ell^2<\frac{3\gamma}{2}$. If $n=0$, the Universe tends to a De Sitter model and the limit $k\rightarrow 0$ is always satisfied.\\
\\
Again, in the Brans-Dicke frame, we deduce that when isotropisation arises and the scalar field vanishes or $n=0$, the metric functions tend to the same form as in the Einstein frame because $\lambda$ tends to a constant. When the scalar field diverges and $n\not =0$, they tend to:
$$t^{\frac{n^2(9\gamma-13)+3(7\gamma-8)\omega_0}{n^2(9\gamma-13)+3\gamma\omega_0}}$$
\begin{figure}[h]
\includegraphics[width=\textwidth]{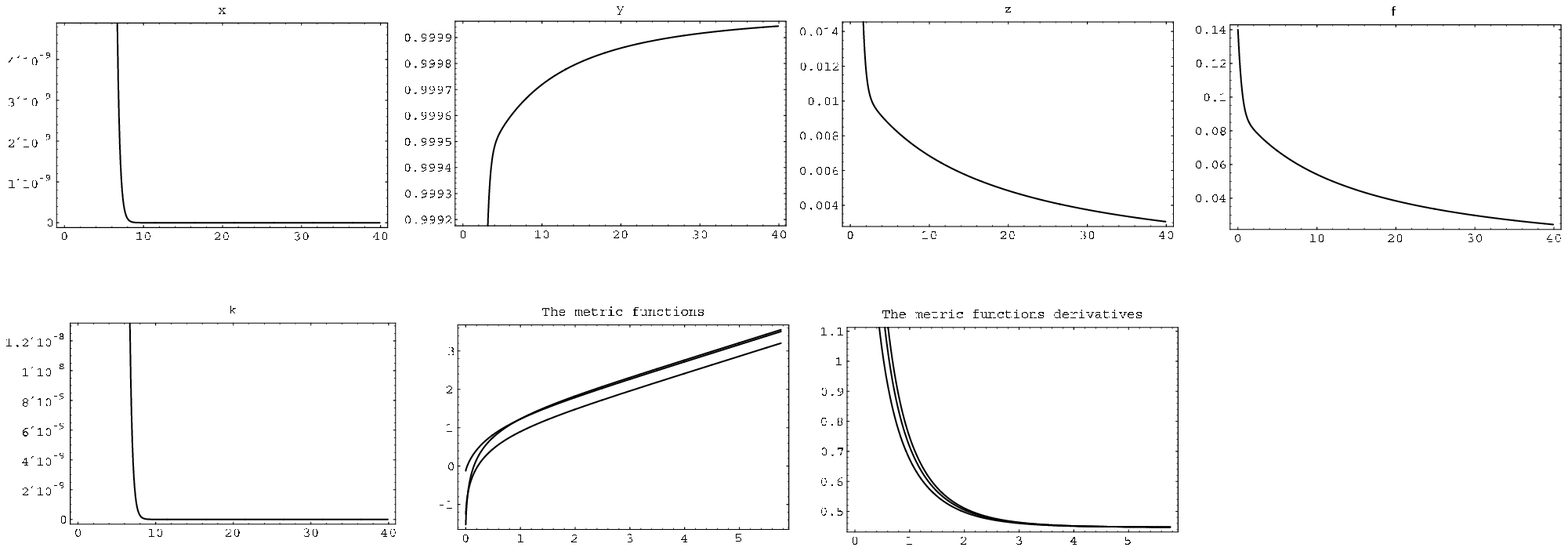}
\caption{\scriptsize{\label{fig7}These figures represent the approach for a class 1 isotropisation when $\omega_0=2.3$, $\omega_1=0.5$, $n=1.5$ and $m=1.1$. $k$ tends to vanish showing that the scalar field energy density dominates the one of the perfect fluid. In the same way, $\phi$ vanishes.}}
\end{figure}
\textbf{\emph{Low energy string theory with a power potential}}\\
We now consider the minimally coupled Lagrangian (\ref{action}) with: 
%-------------------------EQUATION-------------------------------------%
\begin{eqnarray*}
\omega&=&\omega_0\phi^2+\omega_1\\
U&=&\phi^{p} e^{n\phi}\\
\lambda&=&e^{m\phi}
\end{eqnarray*}
Applying the conformal transformation, it is cast into the following non minimally coupled theory:
%-------------------------EQUATION-------------------------------------%
\begin{eqnarray*}
G&=&e^{\frac{m}{3(4-3\gamma)}\phi}\\
\omega&=&\left[\frac{\frac{3}{2}+\omega_0\phi^2+\omega_1}{\phi^{2}}-\frac{3m^2}{18(4-3\gamma)^2}\right]\phi e^{\frac{-m}{3(4-3\gamma)}\phi}\\
U&=&\phi^p e^{(n-\frac{2m}{3(4-3\gamma)})\phi}
\end{eqnarray*}
The law energy string theory with a power potential is recovered when $m=3(4-3\gamma)$, $n=2$, $\omega_0=5/2$ and $\omega_1=-3/2$.\\
Calculating $\ell$ and $\ell_m$, we get:
$$\ell=\frac{p+n\phi}{\sqrt{3+2\phi^2\omega_0+2\omega_1}}$$
$$\ell_m=\frac{m\phi}{\sqrt{3+2\phi^2\omega_0+2\omega_1}}$$
Again, it is impossible that $\ell_m$ diverges and in the same time $\ell<<\ell_m$. Thus the \underline{case 3} is excluded. For the \underline{case 1}, we show that the scalar field behaves as:
$$\phi=p (m-n)^{-1} ProductLog((n-m)e^{3\gamma p^{-1}(\Omega-\phi_0)})$$
When $p\gamma^{-1}>0$, the scalar field vanishes, otherwise it diverges. Then, $(n-m)p^{-1}$ have to be positive otherwise $\phi$ is complex.\\
When $\phi\rightarrow 0$, it is necessary that $3+2\omega_1>0$ such that $\ell$ and $\ell_m$ be real and the reality conditions for the equilibrium points reduce to $2p^2-3\gamma(3+2\omega_1)>0$. Then, the metric functions tend to $t^{\frac{2}{3\gamma}}$ and $\lambda$ to a constant. This case is plotted on figure \ref{fig9}.\\
\begin{figure}[h]
\includegraphics[width=\textwidth]{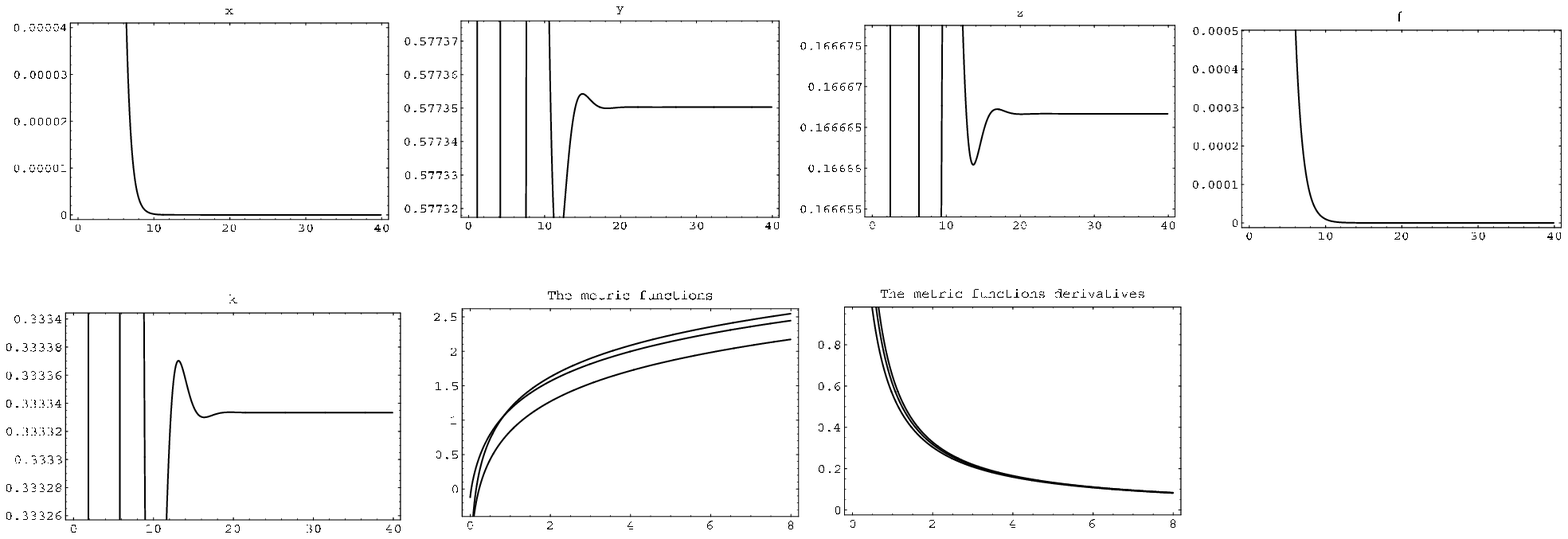}
\caption{\scriptsize{\label{fig9}These figures represent the approach for a class 1 isotropisation when $\omega_0=2.3$, $\omega_1=0.5$, $n=-3.1$, $m=1.1$ and $p=3$. $k$ oscillates to a constant and $\phi$ vanishes. Note the strong oscillations of $y$, $z$ and $k$.}}
\end{figure}
When $\phi\rightarrow \infty$, it is necessary that $\omega_0>0$ such as $\ell$ and $\ell_m$ be real and $n\not = m$ such as $\ell$ does not tend to $\ell_m$. The reality conditions for the equilibrium points write then 
$$2m(m-n)+3\gamma(2-\gamma)\omega_0>0$$
$$n(n-m)-3\gamma\omega_0>0$$
It implies that $n(n-m)>0$ and $n\not =0$. The metric functions tend to $t^{\frac{2(n-m)}{3n\gamma}}$ and $\lambda\rightarrow t^{-2\frac{m}{n}}$. Some figures similar to the figure \ref{fig9} but with diverging $\phi$ may be obtained.\\
\\
In the Brans-Dicke frame, the metric functions tend to the same form as in the Einstein frame during isotropisation if $\phi\rightarrow 0$. When $\phi$ diverges, they tend to:
$$t^{\frac{m(8-5\gamma)+2n(3\gamma-4)}{\gamma\left[m+3n(3\gamma-4)\right]}}$$
Let us examine the \underline{case 2}. The scalar field is such that:
$$\phi_0+1/2\left[\frac{(3+2\omega_1)\ln\phi}{p}-\frac{n^2(3+2\omega_1)+2p^2\omega_0}{pn^2}\ln(p+n\phi)+\frac{2\omega_0\phi}{n}\right]\rightarrow \Omega$$
Hence, it exists three cases such that $\Omega\rightarrow -\infty$.\\
In the first one, $\phi$ tends to vanish and it is then necessary that $p>0$ and $3+2\omega_1>0$. $\ell\rightarrow p(3+2\omega_1)^{-1/2}$ and thus we need $p^2(3+2\omega_1)^{-1}<3$. The metric functions tend to $t^{(3+2\omega_1)/p^2}$. $k$ always tends to 0 as long as $\ell^2<3/2\gamma$. This case is shown on figure \ref{fig8}. Since $\phi$ vanishes, $\lambda$ tends to a constant and the results are the same in the Brans-Dicke frame.\\
\begin{figure}[h]
\includegraphics[width=\textwidth]{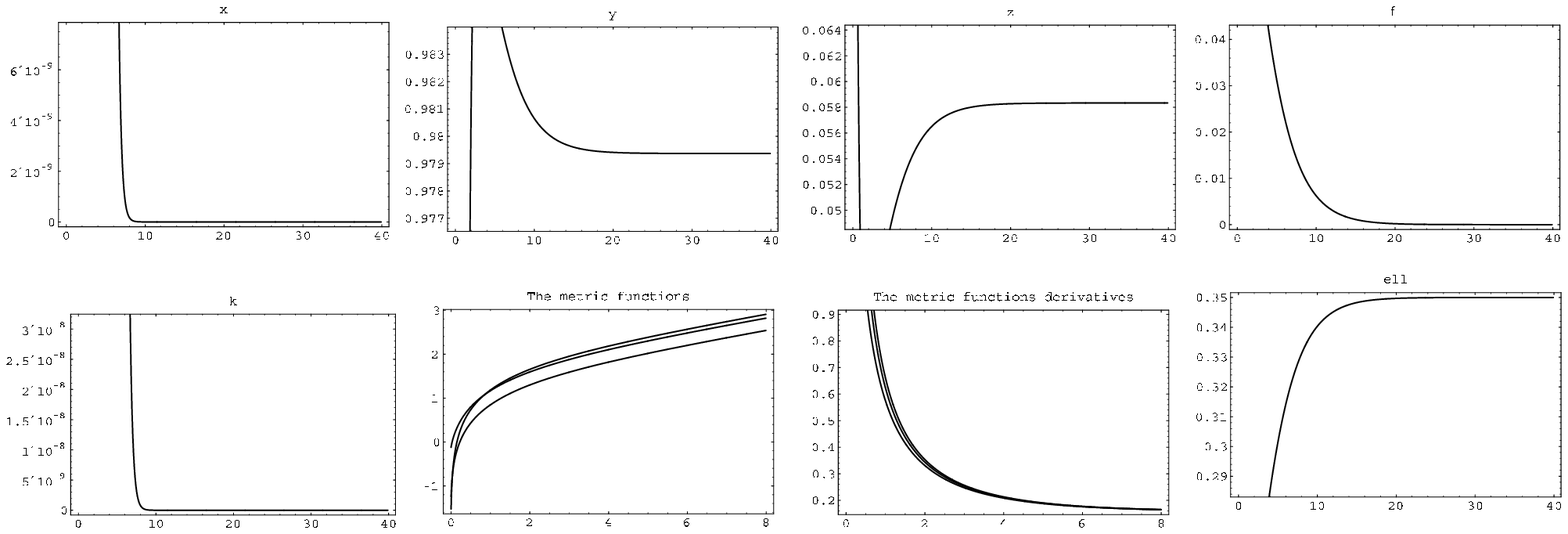}
\caption{\scriptsize{\label{fig8}These figures represent the approach for a class 1 isotropisation when $\omega_0=2.3$, $\omega_1=0.5$, $n=-3.1$, $m=1.1$ and $p=0.7$. $k$ and $\phi$ vanish. $\ell$ tends to $0.35$ which is smaller than $3/2\gamma=3/2$}}
\end{figure}
In the second one, $\phi$ diverges as $\frac{n}{2\omega_0}\Omega$. It must be positive and $\omega_0>0$ thus implying $\phi\rightarrow +\infty$ and $n<0$. Then, $\ell$ tends to $n(2\omega_0)^{-1/2}$ and it follows that a necessary condition for isotropisation is $n^2(2\omega_0)^{-1}>3$. Then, the metric functions tend to $t^{\frac{2\omega_0}{n^2}}$ and $k$ to $0$ if $n(m-2n)+6\gamma\omega_0>0$. In the Brans-Dicke frame, we find that the metric functions tend to $t^{\frac{mn+12\omega_0(3\gamma-4)}{mn}}$.\\
In the third one, $\phi\rightarrow -pn^{-1}$ which implies $\left[-n^2(3+2\omega_1)-2k^2\omega_0\right](pn^2)^{-1}>0$. Then, $\ell\rightarrow 0$ and the Universe tends to a De Sitter model. The condition $k\rightarrow 0$ is always respected. Once again $\lambda$ tends to a constant and in the Brans-Dicke frame, the metric functions tend to the same form as in the Einstein frame.
%------------------------------------------------------------------------------------------------------------------------------------------------------------------------------------------%
\subsection{Conclusion}\label{s33}
We have found some necessary conditions for isotropisation of Bianchi type $I$ model with a massive scalar field, minimally coupled to the curvature but not to the perfect fluid. They depend on the asymptotical behaviours of $k$ and the product $k\ell_m$. We have then deduced the asymptotical behaviours of the metric functions and the potential in the vicinity of the isotropy. A possible solution to the coincidence problem has also been found. Through some applications, we have shown how to extend our results to a scalar-tensor theory. The necessary conditions for isotropisation are the same in the Einstein or Brans-Dicke frames but the asymptotical behaviour of the metric functions are different. They may be determined via a conformal transformation of the metric. We have thus studied the isotropisation of the Brans-Dicke and low energy string theories with some power or exponential laws of the scalar field for the potential.\\
\\
%------------------------------------------------------------------------------------------------------------------------------------------------------------------------------------------%
Parts of the calculus and phase portrait diagrams have been made with help of the marvellous DynPack 10.69 package for Mathematica 4 written by Alfred Clark (http://www.me.rochester.edu/courses/ ME406/webdown/down.html for download).
\appendix
%------------------------------------------------------------------------------------------------------------------------------------------------------------------------------------------%
\section{Perfect fluid conservation law when it is non minimally coupled to the scalar field}\label{a1}
In this appendice, we calculate the energy momentum conservation law of the perfect fluid when it is non minimally coupled to the scalar field. This calculus is also made in \cite{DamNor93} and more particularly in \cite{Bek77}. Let us consider the Lagrangian of a non minimally coupled scalar field also known as hyperextended scalar tensor theory\cite{TorVuc96}:
\begin{equation}\label{lbd}
L=(G^{-1}R-\omega\phi^{-1}\phi_{,\mu}\phi^{,\mu}-U+T^{\alpha\beta}\delta g_{\alpha\beta})\sqrt{g}
\end{equation}
Then we  define a conformal transformation of the metric:
\begin{equation}\label{conforme}
g_{\alpha\beta}=G\bar g_{\alpha\beta}
\end{equation}
%-------------------------EQUATION-------------------------------------%
\begin{equation}\label{tconf}
dt=\sqrt{G}d \bar t
\end{equation}
The frame related to $g_{\alpha\beta}$ is usually called the Brans-Dicke frame whereas the one related to $\bar g_{\alpha\beta}$ is called the Einstein frame. In both cases, $t$ and $\bar t$ are the proper times such as the $00$ metric functions components are $-1$. Applying the transformation (\ref{tconf}) casts the Lagrangian (\ref{lbd}) into:
\begin{equation}\label{lei}
L=\left[\bar R-(3/2(G^{-1})_\phi^2G^{2}+\omega G\phi^{-1})\phi_{,\mu}\phi^{,\mu}-G^2U+G^3T^{\alpha\beta}\delta \bar g_{\alpha\beta}\right]\sqrt{\bar g}
\end{equation}
where the scalar field is now coupled non minimally with the perfect fluid but minimally with the curvature. Consequently, it comes:
\begin{eqnarray*}
\bar T^{\alpha\beta}&=&G^3T^{\alpha\beta}\\
\bar T&=&G^2T
\end{eqnarray*}
We deduce the following energy conservation law:
\begin{eqnarray*}
\bar T^{\alpha\beta}_{;\alpha}&=&3G_{,\alpha}G^2T^{\alpha\beta}\mbox{ (since $T^{\alpha\beta}_{;\alpha}=0$)}\\
\bar T^{\alpha\beta}_{;\alpha}&=&3G_{,\alpha}G^2g^{\alpha\beta}T^\alpha_\alpha\\
\bar T^{\alpha\beta}_{;\alpha}&=&3G_{,\alpha}G^2G^{-1}\bar g^{\alpha\beta}G^{-2}\bar T\\
\bar T^{\alpha\beta}_{;\alpha}&=&3G_{,\alpha}G^{-1}\bar g^{\alpha\beta}\bar T\\
\bar T^{\alpha\beta}_{;\alpha}&=&-3\frac{dG}{dt}G^{-1}\bar T\mbox{ (since $G=G(t)$)}
\end{eqnarray*}
Let us remark that in \cite{Bek77}, this law is interpreted as the action of a force on matter due to the variability of the rest masses. Consequently, matter does not follow the spacetime geodesics. To simplify the calculations, we put $p^*=G^2p$ and $\rho^*=G^2\rho$. Hence, we have $\bar T^{\alpha\beta}=(\rho^*+p^*)u^{\alpha}u^{\beta}+\bar g^{\alpha\beta}p$. Moreover, we have assumed $p=(\gamma-1)\rho$. Thus, it comes:
\begin{eqnarray*}
\bar T^{0\beta}_{;\beta}&=&-3\frac{dG}{dt}G^{-1}(3p^*-\rho^*)\\
\frac{d\rho^*}{dt}+(\rho^*+p^*)V^{-1}\frac{dV}{dt}&=&-3\frac{dG}{dt}G^{-1}(3\gamma-4)\rho^*\\
\rho^{*-1}\frac{d\rho^*}{dt}+\gamma V^{-1}\frac{dV}{dt}&=&-3\frac{dG}{dt}G^{-1}(3\gamma-4)\\
\rho^*V^\gamma=G^{3(4-3\gamma)}
\end{eqnarray*}
From this last result and the expression for the Lagrangian $L_m$ for a perfect fluid calculated in \cite[pages 48-52]{Rya72}, we are able to determine the form of $H_m$, the term describing the matter in the ADM Hamiltonian. Indeed, we have:
\begin{eqnarray*}
L_m&=&T^{\alpha\beta}\delta g_{\alpha\beta}\sqrt{g}\\
&=&-8\pi R_0^3Ne^{-3\Omega}\rho\\
&=&-8\pi R_0^3\bar N e^{-3\bar\Omega}\rho^*\\
&=&-8\pi R_0^3\bar N e^{-3\bar\Omega}G^{3(4-3\gamma)}V^{-\gamma}
\end{eqnarray*}
and consequently:
\begin{equation}
H_m=-24\pi^2\bar g^{1/2}L_m=192\pi^3R_0^3G^{3(4-3\gamma)}e^{3(\gamma-2)\bar\Omega}>0
\end{equation}
We will write symbolically this relation under the form  $H_m=\delta \lambda(\phi)e^{3(\gamma-2)\bar\Omega}$.

\end{document}